\documentstyle [12pt] {article}

\begin{document}
\title {\large \bf  Representations\\ of The Coordinate Ring of
$ GL_{q}(n) $}
\author {Vahid Karimipour}
\date { }
\maketitle
\begin {center}
Institute for studies in Theoretical Physics and Mathematics
\\ P.O.Box 19395-1795 Tehran, Iran \\
{\it  Department of Physics , Sharif University of Technology\\
P.O.Box 11365-9161 Tehran, Iran\\
Email. Vahidka@irearn.bitnet}.
\end{center}
\vspace {10 mm}
\begin {abstract}
{It is shown that the finite dimensional irreducible representations of the
quantum
matrix algebra $ M_q(n) $ ( the coordinate ring of $ GL_q(n) $ ) exist only
when q is a root of unity
( $ q^p  = 1 $ ). The dimensions of these representations can only be one
 of the
following values: $ {p^N \over 2^k } $ where $ N = {n(n-1)\over 2 } $
and $ k
\in \{ 0, 1, 2, . . . N \} $
For each $ k $ the topology of the space of states is $ (S^1)^{\times(N-k)}
\times [ 0 , 1 ] ^{(\times (k)} $ (i.e. an $ N $ dimensional torus for
$ k=0 $
and an $ N $ dimensional cube for $ k = N $ ).}
\end{abstract}
\newpage
\noindent
{\large \bf 1.  Introduction}\\
As far as $ q $ is not a root of unity,there is a well known paralellism
 [1-5] between the representation theory of the universal enveloping
algebras
in
the deformed and the undeformed which results from the paralellism between
the
root space decompositions in these cases.
However for the dual objects (i.e. quantum matrix algebras or the
 coordinate
ring of the algebra of functions on the group ) such a paralellism dose not
exist due to the obvious fact that the classical limit of these objects
is a
free abelian algebra.

In this letter which is an extension of a previous one [6] we pose and
 answer
the following two questions:

1) Is there a natural definition of q-analouge of root systems for the
quantum
matrix algebra $ M_q(n) $ ( the coordinate ring of $ GL_q(n) $ )?.

2) What is the character of the representations of $ M_q(n) $ ?.

In answering the above questions we follow the following line of
 reasoning .\\
We present in section (2), the most essential properties of $M_q(3) $ which
allowed
us to prove in ref.[6] the special $ (n=3) $ case of the results stated in
the abstract of this letter. In section (3) we show that these properties
are also true for $ M_q(n) $ and define a canonical root system for
$ M_q(n) $.
The commutation relations in this root system are very simple and allows us
to characterize various representations of this algebra.
The theorems presented in section (3), are stated without proof. Their
proof coustitute the material of section(4).\\ \\
{\large \bf 2. A resume of Representations of $ GL_q(3) $}\\ \\
Consider a matrix $ T \in GL_q(3) $ \\
$$ T = \left( \begin{array}{lll} a & b & c   \\ d & e & f
\\ g & h & k  \end{array} \right)$$
For the commutation relations between the elements of this matrix see [6].
 The
elements of this matrix generate the quantum algebra $M_q(3) $. The
strategy
introduced in [6], for constructing the representations of $M_q(3)$, was
based on the following observations.\\ \\
{\bf a-} The commuting elements $ c, e  $ and  $ g$  can be simultaneously
diagonalized. They play the
role of Cartan subalgebra generators. We denote the Cartan subalgebra by $
{\bf \Sigma^0 } $\\
{\bf b-} The positive (respectively negative) root systems are generated by
the elements $ f, h $  and  $ \Delta = ek - qfh $ ( respectively, $ b, d $
and
$ \Delta ' = ae-qbd $).  The nice properties of this choice of roots
are that:\\
{\bf b-1} They have multiplicative commutation relations with the
 elements of
${\bf \Sigma^0 }$.
By multiplicative relation between two element $ {\bf x} $ and $ {\bf y} $,
 we
mean a relation of the form $ {\bf x y  } = q^{\alpha }  {\bf y x } $ ,
where $ {\alpha }$ is an integer.\\
{\bf Remark}: In the rest of this paper a multiplicative relation between
${\bf x } $ and $ {\bf y } $is indicated as $ {\bf x y } \approx {\bf y x }
$ \\
{\bf b-2} All the positive (resp. negative) roots commute among themselves.
\\
{\bf b-3} Each negative root has multiplicative relation with all the
positive
roots except one, which we call the positive root corresponding to that
negative root. ( $ f,h $ and $ \Delta $ correspond respectively to $ b,d
$ and
$ \Delta ' $),with commutation relations:
$$ bf - fb = ( q - q^{-1} )ce \hskip 2cm dh - hd = ( q-q^{-1} ) eg $$
$$ q^{-1} \Delta '\Delta - q \Delta \Delta ' = (q^{-1} - q) det_q(T) e $$
\\ We denote the set of positive and negative roots by $
{\bf \Sigma^+ }$ and $ {\bf \Sigma^-} $ respectively.\\
{\bf c-} When q is a root of unity, $ q^p = 1 $ , the p-th power of all the
elements in $ {\bf \Sigma } \equiv  {\bf \Sigma^0 \cup \Sigma^+ \cup
\Sigma^-
 } $ are central.\\
{\bf d-} From the representations of ${\bf\Sigma} $  one can reconstruct
the
representations of $ M_q(n) $ ( i.e. one can recoustruct the
representations
of $a$ and
$k$).\\ The basic strategy is then very simple. One considers a cube of
states\\ $$W = \{ \vert l ,m ,n > = f^l h^m \Delta^n \vert 0 >
\ \ \ \ 0\leq l , m ,
n \leq p-1 \} $$
where $ \vert 0 > $ is a common eigenvector of $ c, e $ and  $ g $
and then shows that $ W $ is invariant under the action of $ {\bf\Sigma }$.
(see [6] for details)\\
Property {\bf b1)} shows that the states of $ W $ are all eigenstates of
$ c,e
$ and
$ g$ .\\ Property {\bf b-2 }) shows that each positive root acts a raising
operator in
one direction of the cube independent of the other positive roots.
(i.e. $f$ in
the $l$ direction, $h$  in the $m$ direction, etc).\\
Property {\bf b-3)} shows that each negative root acts as a lowering
operator only in the same direction of the cube which has been generated by
its corresponding positive root.
Therefore the study of the representations
effectively reduces to the analysis of the action of a typical pair of
positive
and corresponding negative root say (f and b) on a string of states ( i.e.
$ \vert l,m,n> $ for fixed m and n). This analysis is  very much
like the one carried out for the case of $ GL_q(2) $in [7].\\
Property {\bf c)} allows the idendification of opposite sides of the
cube $W$,
independent of each other.\\
These are the basic steps which lead to the classification of $ M_q(3) $
modules. In fact what simplifies the representation theory of $ GL_q(3) $
is the observation that
the q-analouge of non simple roots are the
q-minors of the matrix $ T $ .
Therefore what we need is to prove that the properties ({\bf a - d }) hold
exactly for $ M_q(n) $, this is the subject of the next section
where we define the natural root system of this algebra.\\ \\
{\large \bf 3. The Root system of $ M_q(n) $}

The quantum matrix algebra $ M_q(n) $ is generated by 1 and the elements of
an $ n\times n $ matrix T, subject to the relations [8]:\\
\begin{equation} R \ T_1T_2 = T_2 T_1 R \end {equation}
where R is the solution of the Yang-Baxter equation corresponding to $
SL_q(n) $ [9].
The commutation relations derived from (1) can be neatly expressed in the
following way.\\
For any for elements $ a, b, c, $ and $ d $ in the respective
positions specified by rows and columns (ij) ,(ik),(lj) and (lk),
the following relations hold :
$$ ab =q ba \hskip 1cm cd = q dc  $$
$$ ac =q ca \hskip 1cm bd = q db  $$
$$ bc=cb \hskip 1cm ad-da=(q - q^{-1})bc $$
The quantum determinant of T is also defined as:
\begin{equation} D_q(T) =
\Sigma_{i=1}^{n}(-q)^{i-1}t_{1i}\Delta_{1i}\end{equation}
where $\Delta_{{1i}} $ is the q-minor corresponding to $ t_{1i}$  and is
defined by a similar formula. The q-cofactor of the element $ t_{1i} $
is defined to be $ (-q)^{i-1}\Delta_{1i} $ and is denoted by $ C_{i}$.
In eq. 2 $ D_q(T) $ has been expanded in terms of the elements in the first
row of T . Another useful expansion is in term of the last column of T:
\begin{equation} D_q(T) =
\Sigma_{i=1}^{n}(-q)^{n-i}\Delta_{in} t_{in}\end{equation}

In order to present the multiplicative relations in (2), we devise a
notation which will be also convenient later on, when we will derive the
relations between the q-minors.\\
Any element of the matrix T will be shown by a $\bullet $ and any q-minor of
any size
by a square. The positions of the dots or squares, represent their relative
positions in the matrix T, the order of the elements in a multipliative
relation is shown by an arrow, the factor which is obtained when one
reverses
the sence of arrow is indicated on the arrow. Thus
the multiplicative relations in  (2)
are depicted as follows:\\  \\ \\ \\ \\ \\ \\ \\ Let us label the elements
of
the matrix T as follows:\\
$$ T = \left( \begin{array}{llllllll} .&.&.&.&.& .& Y_1 & H_1
\\.&.&.&.&.&  Y_2 & H_2 & X_1 \\.&.&.&.& Y_3 & H_3 & X_2 & . \\.&.&.&Y_4&
 H_4
&X_3
&.&.\\.&.&.&.&.& .&.&.\\.&.&.&.&.& .&.&.\\Y_{n-1}&H_{n-1}& X_{n-2}
&.&.&.&.&.\\
H_n& X_{n-1} .&.&.&.&.& .
\end{array} \right)$$\\
Consider the elements $H_i ,X_i $ and $ Y_i $ together with the q-minors
(q-determinants of the submatrices)\\
$$ H_{ij} = det_q \left( \begin{array}{llll} .&. &.& H_i \\
.&.&.&. \\
.&.&.&.\\ H_j &.&.&.\end{array} \right)$$
$$ X_{ij} = det_q \left( \begin{array}{llll} .&. &.& X_i \\
.&.&.&. \\ .&.&.&.\\ X_j &.&.&.\end{array} \right)$$
$$ Y_{ij} = det_q \left( \begin{array}{llll} .&. &.& Y_i \\
.&.&.&. \\ .&.&.&.\\ Y_j &.&.&.\end{array} \right)$$
(i.e. $ H_{12}= det_q \left(\begin{array}{ll}Y_1&H_1 \\H_2& X_1
\end{array}\right) = Y_1X_1-q H_1 H_2 )$
For convenience we sometimes denote $H_i,X_i $ and $ Y_i $ by $ H_{ii},
X_{ii}
$ and $ Y_{ii} $ respectively.\\
Our choice for the Cartan subalgebra, and the positive and negative root
system is as follows. These subalgebras are generated respectively by the
elements $ H_{i} , X_{ij} $ and $ Y_{ij} $. We denote these subalgebras
by $ {\bf \Sigma^0 ,\Sigma^+} $ and ${\bf \Sigma_-} $ respectively.
This choice is rather unique because of the following propositions.\\ \\
{\bf proposition 1}\\
$$[H_{ij},H_{kl}] = 0 $$
$$[X_{ij},X_{kl}] = 0 $$
$$[Y_{ij},Y_{kl}] = 0 $$\\
{\bf note}: the elements $ H_{ij}$ do not belong to $\Sigma^0$. We have
included their relation here for later use.\\ \\
{\bf proposition 2}\\
$$ H_i X_{ij} = q X_{ij}H_i \hskip 1cm \forall j $$
$$ H_{j+1} X_{ij} = q X_{ij}H_{j+1} \hskip 1cm \forall i $$
$$ H_k X_{ij} =  X_{ij}H_k \hskip 1cm  k \ne i , j+1 $$
with $ (q\longrightarrow q^{-1}  , X_{ij}\longrightarrow Y_{ij} ) $\\
 {\bf proposition 3}.\\
 $$ H_{ij} X_{kl} \approx  X_{kl}H_{ij} \hskip 1cm
X\longrightarrow Y $$
$$ Y_{kl} X_{ij} \approx X_{ij}Y_{kl} \hskip 1cm
(k,l)\ne(i,j)$$
$$ Y_{i} X_{i} - X_{i}Y_{i} = (q - q^{-1} ) H_{i}H_{i+1}$$
$$ q^{-1} Y_{ij} X_{ij} - q X_{ij}Y_{ij} = (q^{-1}-q) H_{i,j+1}H_{i+1,
j}$$\\
{\bf proposition 4}. For $ q^p=1$ the p-th power of all the elements of
$\Sigma$ are central.\\

Let V be a $ {\bf\Sigma}$ module. We call this module trivial if,the
 action of
one or more of
the elements of ${\bf \Sigma}$ on it, is identically zero. We are
interested in
nontrivial
$ {\bf\Sigma}$ -modules. (the trivial one's are representations of
reductions
of ${\bf \Sigma}$ ).\\ \\
{\bf proposition 5}. A $ {\bf\Sigma}$ module V is nontrivial only if the
following condition holds. all the subspaces
$$ K_{ij}\equiv \{ v\in V \vert H_{ij}v = 0 \}$$ must be zero
dimensional.\\ \\
{\bf Proof}. The proof is exactly paralell to the case of $M_q(3)$, and is
based on the multiplicativity of $ H_{ij} $'s  with all the elements of $
{\bf \Sigma } $.\\ \\
{\bf proposition 6}:i) Finite dimensional irreducible representations
 of $ {\bf
\Sigma } $exist only when q is a root of unity.\\ ii) Any non-trivial ${\bf
\Sigma }$ module V is also an $ M_q(n) $  module and vice versa.\\ \\
{\bf proof}: The proof of this proposition is exactly parallel to the
case of
$ M_q(3)$. One uses the expressions (2)(resp. 3) for the q-determinants
$ Y_{ij} $(resp. $ X_{ij}$ (starting
from $ j=i+1 $,continuing to $ j=i+2,i+3 ... $) and uses the fact that
in the representation of $ {\bf \Sigma}$, all the
elements $ H_{ij} $ are invertible diagonal matrices. Note that
invertibility of $H_{ij}$'s
(due to proposition. 5) is crutial here, otherwise one can not define the
actions of the remaining elements of T or V.\\ \\
At this stage one considers a hypercube of states W:
$$ W := \{ \vert {\bf l >} = \prod_{i=1}^{N} E_i^{l_i}\vert 0 >,\hskip
1cm  E_i
\in {\bf \Sigma^+} \hskip 1cm , 0\leq l_i \leq p-1 \} $$
where  $ \vert 0 > $ is a
common eigenstate of $ H_{i},s $ and shows that W is invariant under
the action
of ${\bf \Sigma} $.
Denote the negative root corresponding to $ E_i $ by $ F_i $ and let
the values of their p-th power on W be respectively the numbers
$ {\eta_i}^+ $
and $ {\eta_i} ^- $ .
Depending on the values of these parameters this pair of roots
either traverse a full p-dimensional cycle
( when ${ \eta_i} ^+ \ne 0 ,  {\eta_i} ^- \ne 0 )  $ or traverse a
$ {p\over 2
} $ dimensional line segment with highest and lowest wieght ( when
$ {\eta_i} ^+ = {\eta_i }^- = 0 )  $ in the i-th direction of the cube .
An intermediate case also occurs when only one of the parameters is zero in
which case the representation is semicyclic in that direction.
The detailed analysis
is a word by word repeatition of that carried out in [6] for $ GL_q(3) $.
This proves our statement made in the abstract of this letter.\\ \\
We now proceed to the next section where the general method for deriving
the
the commutation relations in
$ {\bf \Sigma } $ are presented.\\ \\ \\
{\large \bf 5. Commutation relations in {\bf $ \Sigma$}}\\

We use the graphical notation introduced in section 4.
The first basic fact is presented in the
following lemma.\\ \\ {\bf Lemma 7}. \\ \\ \\ \\
{\bf Proof}: Expand the determinant and note that all of the
relations are of type (4-c) except the ones on the lower edge, which is of
type (4-a).\\
We combine diagram 1 with three similar relations in the following
 diagram.\\
\\ \\ \\ \\   {\bf Lemma 8}.\\ \\ \\ \\
 {\bf Proof: } Let the minor be $ n\times n $ . For $ n=2 $ direct
calculation
verifies the statement. We use induction on  $ n $. consider fig.(1).
Writing $ \Delta_{n+1} $ as $$ \Delta_{n+1} = \Sigma d_i C_i $$ where
$ C_i $
is the cofactor of  $ d_i $ in $ \Delta_{n+1} $  and passing a through
$ d_i
$,we have: $$ a\Delta_{n+1} = \Sigma \Big( d_i a + (q-q^{-1}\Big) b c_i)
C_i$$
We now use the assumption of induction ( $ a C_i = q C_i a $ )
and the property of the determinant ( $ \Sigma a_i C_i = 0 $ ) to arrive at
the final result.

It's combined with three other relations in the following diagram.\\ \\
\\ \\ \\ {\bf Lemma. 9}\\ \\ \\ \\
{\bf Proof}: Write the q-minor as $ \hskip 2cm $ which symbolically means
that
the q-minor is the sum of the products of the elemsnts of A and
 q-cofactors in
B. Passing the $ \bullet $ through A gives the factor 1 and passing it
through
B gives q. \\ \\
{\bf Coralarry}: {\bf a) }  \\ \\ $$ \hskip 6cm
\Delta \Delta ' =  q^l \Delta ' \Delta  $$   \\ \\

{\bf Proof}: expand the left hand minor and use lemma (9).\\ \\
The $\bullet $  ( resp. the small minor) can be in other similar
positions as in the previous two lemmas, with appropriate factors of $q$
or $
q^{-1} $ ( resp. $ q^l $ or $ q^{-l} $ )\\ \\
 {\bf Lemma 10 : } \\ \\ \\ \\ \\
 {\bf Proof: } Consider fig. (2). We use induction on $ l'$. For $ l' = 0 $
the result is true. Assume that its true for $ l' $ and  write $ \Delta$
as : $
\Delta = \Sigma _{i=1}^{m} a_i C_i $
where $ C_i $ is the q-cofactor of $ a_i $ in $ \Delta$ and use the
results of
the previous lemmas:i.e:
$$ a_i \Delta' = \Delta' a_i \hskip 1cm C_i \Delta' = q^{-l'}q^{l-1}
\Delta' c_i
\hskip 2cm 1\leq i < l $$
$$ a_i \Delta' = q^{-1}\Delta' a_i \hskip 1cm C_i \Delta' =
q^{-l'}q^{l}\Delta' c_i \hskip 2cm l\leq i \leq m $$
from which we obtain the result for $ l'+1 $.\\ \\
 {\bf Important Remark}:\\
In the matrix T, there are many more positions of q-minors which give
 rise to
very complicated commutation relations. But in ${\bf \Sigma }$ there is
none
(as
the reader can verify ) other than those between $X_{ij}$ and $Y_{ij}$,
which
we now compute exactly.\\ \\
{\bf Proof of the last relation in proposition 3. }\\
Consider fig. (3).
Write $ H_{i,j+1} $  as  $ H_{i,j+1} = a_1 C_1 +  a_2 C_2+
... $ where $ C_{1}= X_{ij} $ is the q-cofactor of  $a_1 $ in the big
matrix. $ H_{i,j+1} $, being the determinant of the big matrix, commutes
with $
a_1$. On the other hand:
$$ a_1 H_{i,j+1} = a_1\big( a_1 X_{ij} + \Sigma_{i\geq 2 }a_i C_i \big)
$$
We now use the fact that for $ i\geq 2 \ \ ,  a_1 a_i = q
a_i a_1 , a_1 C_i = q C_i a_1 $ (lemma 8) and pass $ a_1 $ through
$ \Sigma
a_i C_i $ to find: $$ a_1 H_{i,j+1} = a_1^2 X_{ij} + q^2 \big( H_{i,j+1} -
a_1 X_{ij} \big)a_1 $$
Multiplying both sides from the left by $ a^{-1} $ we obtain:
\begin{equation} a_1 X_{ij} - q^2 X_{ij} a_1 = ( 1-q^2 ) H_{i,j+1} \end
{equation}
(Note: direct calculations which do not need invertibility of $ a_1 $
confirm
this equation ). Now we expand $ Y_{ij} $ in :
$$ Y_{ij} X_{ij} = \big ( a_1 \hat C_1 +\Sigma _{k\geq 2 }  a_k \hat
C_k\big)X_{ij} $$ where
$\hat C_k$ is the cofactor of $ a_k $ in the matrix $ Y_{ij} $.
Note that: $ \hat C_1 = H_{i+1,j} $.\\
{}From lemmas (8-9) we have for $ k\geq 2 $: $$ \big(a_k \hat C_k\big) X_{ij}
= q^2 X_{ij} \big(a_k \hat C_k\big) $$
Therefore $$ Y_{ij}X_{ij} = a_1  X_{ij} H_{i+1,j} + q^2 X_{ij} \big(Y_{ij}-
a_1
H_{i+1,j} \big) $$ Combining this with eq.(5) we obtain the final
 result: i.e.
$$ q^{-1}Y_{ij} X_{ij} - q X_{ij}Y_{ij} =  (q^{-1}-q )H_{i,j+1}H_{i+1
,j}$$ \\ \\ \\
 {\large \bf Acknowledgement:} I would like to thank M.Khorrami and S.
Shariati
for valuable comments.
\newpage
{\large \bf References}
\begin{enumerate}
\item  G. Lusztig , Adv. Math. 70, 237 (1988); Contemp. Math.
82,59 (1989)
\item  M. Rosso , Commun. Math. Phsy. 117, 581 (1988) ; 124, 307 (1989)
\item  R. P. Roche and D. Arnaudon , Lett. Math. Phsy. 17, 295 (1989)
\item  C. De Concini and V. G. Kac , Preprint (1990)
\item  P. Sun and M. L. Ge , J. Phys. A 24, 3731 (1991)
\item  V. Karimipour ,Representations of the coordinate ring of $ GL_q(3)$;
Lett. Math. Phys. (1993) to appear.
\item  M. L. Ge, X. F. Liu , and C. P. Sun : J. Math. Phys. 38 (7) 1992
\item  N. Reshetikhin , L. Takhtajan , and L. Faddeev ; Alg. Anal. 1, 1
78 ( 1989) in Russian
\item  M. Jimbo ; Lett. Math. Phsy. 10, 63 ( 1985) ; 11,247 (1986)
\end{enumerate}
\vfil\break
\newpage
\end{document}